\documentclass[twocolumn]{emulateapj}
\usepackage{multirow}
\usepackage{graphicx}
\usepackage{footnote}
\usepackage{natbib}
\usepackage[fleqn]{amsmath}
\usepackage{color}

\shorttitle{Optimized radio follow up of BNS mergers}
\shortauthors{Carbone \& Corsi}

\begin{document}

\title{Optimized radio follow up of binary neutron-star mergers}

\author{Dario Carbone\altaffilmark{1,$\star$}, Alessandra Corsi\altaffilmark{1}}
\altaffiltext{1}{Department of Physics and Astronomy, Texas Tech University, Box 1051, Lubbock, TX 79409-1051, USA}
\altaffiltext{$\star$}{Email: \email{dario.carbone@ttu.edu}}

\begin{abstract}
Motivated by the recent discovery of the binary neutron-star (BNS) merger GW170817, we determine the optimal observational setup for
detecting and characterizing
radio counterparts of nearby ($d_L\sim40$\,Mpc) BNS mergers.
We simulate GW170817-like radio transients, and
radio afterglows generated by fast jets with isotropic energy $E_{\rm iso}\sim 10^{50}$\,erg, expanding in a low-density interstellar medium
(ISM; $n_{\rm ISM}=10^{-4}-10^{-2}$\,cm$^{-3}$), observed from different viewing angles (from slightly off-axis to largely off-axis). We then
determine the optimal timing of GHz radio observations following the precise localization of the BNS radio counterpart candidate, assuming
a sensitivity comparable to that of the  Karl G. Jansky Very Large Array. The optimization is done so as to ensure that properties such
as viewing angle and circumstellar density can be correctly reconstructed with the minimum number of observations.
We show that radio is the optimal band to explore the fastest ejecta from BNSs in low-density ISM, since the optical emission is likely
to be dominated by the so-called ``kilonova'' component, while X-rays from the jet are detectable only for a small subset of
the BNS models considered here. Finally, we discuss how future radio arrays like the next generation VLA (ngVLA) would improve the detectability
of BNS mergers with physical parameters similar to the ones here explored.
\end{abstract}


\section{Introduction} \label{sec:intro}
On August 17$^{\rm th}$ 2017, a merger of two neutron stars (NSs) was observed for the first time by the LIGO and Virgo gravitational wave
(GW) detectors  \citep{Abbott2017}. A short $\gamma$-ray burst (SGRB) was detected by the \textit{Fermi} and \textit{Integral} satellites only
$\approx 2$\,s after the merger  \citep{Abbott2017b}, followed by the optical/IR/UV signature of a kilonova, and finally by an X-ray and radio
afterglow \citep[][and references therein]{Abbott2017c}. While confirming the predicted link between SGRBs and BNS mergers
\citep{Eichler1989,Narayan1992}, GW170817 also showed some remarkable differences with respect to the previously known
population of cosmological SGRBs \citep[e.g.,][]{Fong2017}. The host galaxy of GW170817, NGC~4993, is located at a distance of about
40\,Mpc \citep{Coulter2017}, making GW170817 the closest SGRB observed to date. GW170817 was sub-energetic in its $\gamma$-ray
emission. Its optical counterpart was dominated by a kilonova component in the first week after the merger
\citep[e.g.,][]{Coulter2017,Cowperthwaite2017,Kasen2017,Smartt2017,Soares2017,Pian2017,Tanvir2017,Valenti2017}, and thus was much
brighter than a short GRB optical afterglow. The delayed onset of its radio-to-X-ray afterglow
\citep[e.g.,][]{Evans2017,Hallinan2017,Haggard2017,Margutti2017,Troja2017} suggested that, if a relativistic jet accompanied GW170817,
then it likely was $\gtrsim 20-30$\,deg off-axis \citep[e.g.,][]{Kasliwal2017,Lazzati2017a,Lazzati2017}.

While the presence of fast jets that successfully break out from the BNS ejecta is predicted by models positing BNS mergers as central
engines of SGRBs, the relatively slow rise of GW170817 radio afterglow requires the presence of emission coming from a wide-angle outflow
\citep{NakarPiran2018,Mooley2017,Lazzati2017}. The last could be related to a structured jet
\citep[i.e., a jet with a relativistic core and slower wings, different from the uniform jets usually invoked to model cosmological SGRBs;
e.g.,][]{vanEerten2012,Lazzati2017a,Lazzati2017},
and/or to a quasi-spherical, mildly relativistic outflow \citep[such as the high velocity tail of the neutron-rich
dynamical ejecta, or a cocoon;][]{Kasliwal2017,Mooley2017}. The turn over of the radio and X-ray light curves of GW170817 at
$t\gtrsim\,150$\,d \citep{Dobie2018,Alexander2018} can also be explained by both an afterglow from a structured successful jet, or by a cocoon
associated with a chocked jet that was collimated before being choked \citep{NakarPiran2018}. Thus, the questions of how common is
GW170817-like emission in BNS mergers, and are there any intrinsic differences (beyond viewing angle ones) between cosmological SGRBs
and GW170817, remain open. Answering them requires building a larger sample of GW-triggered BNS mergers with radio afterglows.

Motivated by the above considerations, here we present a study aimed at optimizing strategies for detecting radio counterparts of BNS
mergers. We work under the reasonable assumption that the isotropic optical kilonova emission will enable precise localizations of the closest BNSs,
and prompt follow-up with sensitive radio arrays with small fields of view \citep[like the Karl G. Jansky Very Large Array - JVLA;][]{Perley2011}.
We include in our study: (i) radio emission from uniform off-axis relativistic jets simulated using the BOXFIT v2 code  \citep{vanEerten2012}
and with model parameter values drawn from those suggested for GW170817
\citep[see e.g.,][]{Alexander2017,Kim2017,Hallinan2017,Haggard2017,Granot2017,Murguia-Berthier2017,Troja2017,Margutti2018};
and (ii) GW170817-like radio transients where wider-angle ejecta contribute to the radio emission. For these, we use the structured jet light
curves described in \citet{Lazzati2017}.

This paper is organized as follows. In Section~\ref{sec:methods} we describe in detail our simulations; in Section~\ref{sec:results} we present
our major results; in Section~\ref{sec:discussion} we compare expectations in the radio band with those in optical and X-rays; finally, in
Section~\ref{sec:conclusions} we conclude.

\section{Method}\label{sec:methods}
\begin{figure}
\begin{center}
\includegraphics[width=0.49\textwidth,viewport=2 0 530 400,clip]{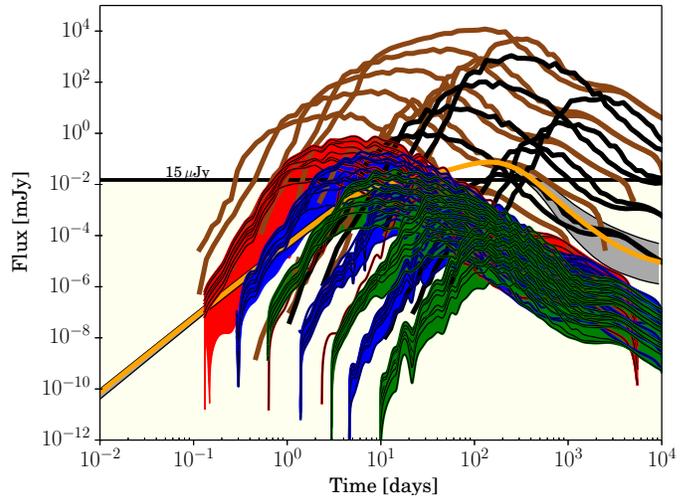}
\caption{Target and contaminant radio (5\,GHz) light curves used in this study (assuming $d_L=40$\,Mpc).
Light curves of our target sources are plotted in red, blue, and green (off-axis SGRBs with n$_{ISM}$\,=\,10$^{-2}$, 10$^{-3}$, and
10$^{-4}$\,cm$^{-3}$ respectively) and orange (GW170817).
Light curves of contaminant sources are plotted in: brown for off-axis LGRBs; black for off-axis HL-SGRBs.
The parameters of all model light curves are given in Table~\ref{tab:models}.
See text in Section~\ref{sec:methods} for more details on these light curves.}
\label{fig:models}
\end{center}
\end{figure}

To establish the optimal observational strategy for detecting and identifying the physical properties of radio counterparts of BNS mergers, 
we simulate off-axis short SGRB and GW170817-like radio light curves (hereafter also referred to as ``target'' sources; Figure~\ref{fig:models}
and Section~\ref{sec:models}).

Based on past experience \citep[e.g.,][]{Kasliwal2016,Stalder2017}, we make the reasonable assumption that optical observations will provide
accurate localizations and distance measurements (via host galaxy identification) of one or several candidate optical counterparts by mapping
the GW error region. For nearby, highly-significant, and well-localized events like GW170817, the true optical counterpart may be unambiguously
identified $\lesssim 24$\,hrs after the LIGO-Virgo trigger \citep{Abbott2017c}, thus enabling prompt follow-up in the radio free of any contamination.
For events with larger localization areas, several optical transients may initially be followed-up in the radio in search for the true counterpart to the
GW trigger \citep[e.g.,][]{Palliyaguru2016,Bhalerao2017,Corsi2017}.

In light of the above considerations, we also simulate ``contaminant'' sources (Figure~\ref{fig:models}; see Section~\ref{sec:models} for
more details). These are radio-emitting transients whose radio light curves may, over one or more epochs, look similar to that of radio
counterparts of GWs.
As such, these contaminants may lead to misidentification of the binary merger electromagnetic counterparts and may introduce degeneracy in the
determination of the physical parameters of BNS fast ejecta
when only a limited number of radio follow-up observations are performed.
We include as contaminants what we refer to as off-axis long GRBs \citep[LGRBs;][]{vanEerten2012}, that are related to fast ejecta expanding
in an ISM of density higher than typically expected for SGRBs and/or GW170817-like events; and other transients dubbed higher-luminosity
short GRBs \citep[HL-SGRBs;][]{vanEerten2012}, that are fast ejecta expanding in a low-density ISM but with energies higher than expected for
SGRBs/GW170817-like events.

As we describe in more detail in what follows, we simulate 4000 realizations of observations of each target model by randomizing the time of the
first radio observation ($t_{\rm radio,0}$), and the flux measured at that and following observational epochs. We then determine the minimum
number of follow-up observations (and their corresponding epochs) required to maximize (globally, among the various possible targets) the
probability of uniquely associating the observed fluxes with the correct targets (i.e., with sources having the correct physical properties, as
discussed in Section~\ref{sec:models}) when observations are compared with our bank of light curve models (Table~\ref{tab:models}).
We set a maximum of ten on the total number of observations that can be performed for each event.
This is a reasonable assumption for a typical JVLA time allocation, considering that each epoch in our simulations consists of a 2\,hr-long observation.
We report results for our optimized radio follow-up strategy both with and without inclusion of contaminants, to address both GW170817-like cases of
well localized GW triggers, and that of less optimal follow-ups where more than one optical counterpart candidate may reported in the GW error region. 

\begin{table}
\begin{center}
\caption{Summary of light curve models used in this work.
Each of the target families reported here consists of nine different models obtained combining $\epsilon_E$ values of 0.1, 0.05, and 0.01 and
$\epsilon_B$ values of 0.01, 0.005, and 0.001.
See text for description.
}
\label{tab:models}
\begin{tabular}{lllll}
\hline
\hline
      &Type &	$E_{\rm iso}$	&	$n_{\rm ISM}$	&	$\theta_v$ \\
	&	& (erg)	&	(cm$^{-3}$)	&	(deg) \\
\hline
Target 0 & SGRB & 10$^{50}$	&	10$^{-2}$	&	20	\\
Target 1 & SGRB & 10$^{50}$	&	10$^{-2}$	&	30	\\
Target 2 & SGRB & 10$^{50}$	&	10$^{-2}$	&	45	\\
Target 3 & SGRB & 10$^{50}$	&	10$^{-3}$	&	20	\\
Target 4 & SGRB & 10$^{50}$	&	10$^{-3}$	&	30	\\
Target 5 & SGRB & 10$^{50}$	&	10$^{-3}$	&	45	\\
Target 6 & SGRB & 10$^{50}$	&	10$^{-4}$	&	20	\\
Target 7 & SGRB & 10$^{50}$	&	10$^{-4}$	&	30	\\
Target 8 & SGRB & 10$^{50}$	&	10$^{-4}$	&	45	\\
GW170817 & - & - & - & - \\
\hline
Contaminant & HL-SGRB & 10$^{51}$	&	10$^{-3}$	&	23	\\
Contaminant & HL-SGRB & 10$^{51}$	&	10$^{-3}$	&	45	\\
Contaminant & HL-SGRB & 10$^{52}$	&	10$^{-3}$	&	23	\\
Contaminant & HL-SGRB & 10$^{52}$	&	10$^{-3}$	&	45	\\
Contaminant & HL-SGRB & 10$^{53}$	&	10$^{-3}$	&	23	\\
Contaminant & HL-SGRB & 10$^{53}$	&	10$^{-3}$	&	45	\\
Contaminant & HL-SGRB & 10$^{54}$	&	10$^{-3}$	&	23	\\
Contaminant & HL-SGRB & 10$^{54}$	&	10$^{-3}$	&	45	\\
\hline
Contaminant & LGRB & 10$^{50}$	&	1	&	23	\\
Contaminant & LGRB & 10$^{50}$	&	1	&	45	\\
Contaminant & LGRB & 10$^{51}$	&	1	&	23	\\
Contaminant & LGRB & 10$^{51}$	&	1	&	45	\\
Contaminant & LGRB & 10$^{52}$	&	1	&	23	\\
Contaminant & LGRB & 10$^{52}$	&	1	&	45	\\
Contaminant & LGRB & 10$^{53}$	&	1	&	23	\\
Contaminant & LGRB & 10$^{53}$	&	1	&	45	\\
Contaminant & LGRB & 10$^{54}$	&	1	&	23	\\
Contaminant & LGRB & 10$^{54}$	&	1	&	45	\\
\hline
\end{tabular}
\end{center}
\end{table}

\subsection{Model radio light curves}\label{sec:models}
\subsubsection{Off-axis uniform relativistic jets} \label{sec:targets}
To simulate 5\,GHz radio light curves of off-axis SGRB targets and LGRB/HL-SGRB contaminants, we use BOXFIT v2. 
The BOXFIT light curves depend on several parameters: the luminosity distance ($d_L$); the jet half-opening angle ($\theta_j$); the viewing
angle ($\theta_v$); the total explosion energy  ($E_{\rm iso}$); the interstellar medium density ($n_{\rm ISM}$); the power-law index of the
shocked electrons distribution $(p)$; and the fraction of the energy converted into magnetic and electric fields ($\epsilon_B$ and $\epsilon_E$). 

When creating our target SGRB models, we fix some of these parameters to simulate GW170817-like events in terms of energetics, distance,
and spectral index. Specifically, we set $d_L=40$\,Mpc \citep{Abbott2017c}, $E_{\rm iso}=10^{50}$\,erg \citep{Granot2017,Hallinan2017,Troja2017},
and $p=2.5$ \citep{Hallinan2017,Mooley2017,Dobie2018}. We also set $\theta_j$\,=\,12\,deg as a reasonable value based on cosmological GRBs
observations and early GW170817 radio light curve models \citep[e.g.,][]{Berger2014,Hallinan2017} although, as we explain below, the specific value
of $\theta_j$ is not crucial as we span a range of viewing angles $\theta_v$.

The remaining model parameters are varied so as to simulate different viewing geometries and BNS merger environments, and to account for
current uncertainties in the microphysics of relativistic shocks. More specifically, we vary $n_{\rm ISM}$ between $10^{-2}$ and
$10^{-4}$\,cm$^{-3}$ \citep[][]{Fong2015,Granot2017,Hallinan2017}, $\epsilon_B$ between $10^{-2}$
and $10^{-3}$, $\epsilon_E$ between $10^{-1}$ and $10^{-2}$ \citep{Santana2014}, and $\theta_v$ between 20\,deg and 45\,deg (from slightly to
largely off-axis). We note that sources with $\theta_v\gtrsim 45$\,deg are not detectable even with the highest circumstellar density in the range
we tested. We also note that because we span a large range of $\theta_v$ values, the specific value of $\theta_j=12$\,deg is not crucial as what
matters observationally is the difference between $\theta_j$ and $\theta_v$.  The resulting light curves are plotted in red, blue and green
(for $n_{\rm ISM}$ of $10^{-2}$\,cm$^{-3}$, $10^{-3}$\,cm$^{-3}$, and $10^{-4}$\,cm$^{-3}$ respectively) in Figure~\ref{fig:models} (where we
neglect redshift corrections).

For off-axis LGRB contaminants (see Section~\ref{sec:methods}) we use the same BOXFIT v2 models where we set $d_L=40$\,Mpc,
$n_{\rm ISM}=1$\,cm$^{-3}$ (which is the average value found in LGRBs), $E_{\rm iso}\,=\,10^{50}-10^{54}$\,erg, and $\theta_v$\,=\,23-45\,deg
(slightly and largely off-axis). The resulting light curves are plotted in brown in Figure~\ref{fig:models}.

Finally, for HL-SGRB contaminants we set $d_L=40$\,Mpc, $E_{\rm iso}\,=\,10^{51}-10^{54}$\,erg, $\theta_v\,=\,23-45$\,deg, and
$n_{\rm ISM}\,=\,10^{-3}$\,cm$^{-3}$ (these are similar to what used for target sources, with the exception of $E_{\rm iso}$ which is now higher;
see the black curves in Figure~\ref{fig:models}). We note that HL-SGRBs with viewing angle of 23\,deg have radio light curves very similar to the
ones from LGRBs with viewing angle of 45\,deg. This highlights the degeneracy between various source parameters, $n_{ISM}$ and $\theta_{v}$
in this case. The resulting light curves are plotted in  black in Figure~\ref{fig:models}.

\subsubsection{GW170817-like transients}
We include in our targets also GW170817-like transients that we simulate using as template the best fit structured jet model by \citet{Lazzati2017}.
To account for possible uncertainties, we consider this model's predictions for the radio flux at each epoch affected by an error that maps the
3$\sigma$ range around the best fit as provided by \citet{Lazzati2017}. The light curve of GW170817 is plotted in orange in Figure~\ref{fig:models},
while the $3\sigma$ envelope around it is displayed by a gray shaded region.

\subsection{Monte Carlo simulations}
For each of the target models, we generate 4000 observed light curves drawing from Gaussian distributions with mean equal to the model flux
at each epoch, and standard deviation equal to the noise fluctuations in the observed flux.
All observations are assumed to be carried out with an RMS noise level of $5\,\mu$Jy. This simulates observations with the JVLA in its most
compact configuration (which provides a conservative estimate of the sensitivity) for a total observing time (including overhead) of 2\,hrs per
epoch, with $\approx$ 15\% bandwidth loss on a nominal 4\,GHz bandwidth (3 bit) due to RFI. At any epoch when the simulated model flux
is below our detection threshold of $3\times$\,RMS, the measured flux is set to zero and the error on it is set equal to the noise RMS.

\subsection{Optimizing the radio follow-up campaign} \label{sec:urgency}
We explore observational strategies in which the first observation in radio is always carried out as soon as possible, at
$t_{\rm radio,0}=t_{\rm opt,0}+\Delta t_{\rm urgency}$
(having assumed $t\,=\,0$ as the time of the merger). Here $t_{\rm opt,0}$ accounts for a possible delay between the merger and the time at
which an accurate localization is provided via identification of an optical counterpart. We randomize $t_{\rm opt,0}$ uniformly in the range
$1-3$\,d. With only one BNS event (GW170817) studied so far, the choice of $t_{\rm opt,0}$ is somewhat arbitrary. However, it is not unreasonable
to assume that the unambiguous identification of a kilonova, including spectral classification, precise localization, and distance determination,
may take up to $\approx 1$\,d (e.g. due to weather or other day time constraints). This lower bound on $t_{\rm opt,0}$ is somewhat conservative
compared to the actual time line of GW170817 optical follow-up. On the other hand, because we expect that several optical facilities will
participate in the follow-up of future LIGO-Virgo BNS triggers, we consider it unlikely that an optical counterpart would be identified later
than $\approx 3$\,d since merger (including weather / day time constraints, and constraints related to the availability of spectroscopic
facilities). $\Delta t_{\rm urgency}$ allows for a possible further delay between the optical identification and the earliest radio observation.
We tested three different ranges: $\Delta t_{\rm urgency}=1\,{\rm hr}-2$\,d (hereafter dubbed high-urgency follow up),
$\Delta t_{\rm urgency}=3-5$\,d (hereafter referred to as medium-urgency follow up), and $\Delta t_{\rm urgency}=7-15$\,d (low urgency).

The ultimate goal of our simulations is to determine the minimum number of radio follow-up observations, $n_{\rm min}$ (where 1$\le n\le$10),
and their corresponding epochs $\Delta T_n=\,t_{\rm n}\,-\,t_{\rm radio,0}=\,M_n\times 2$\,d (where $M_n$ is an integer in the range
$0\le M_n \le 183$ for a typical 1\,yr-long JVLA observing program, and $\Delta T_0=M_0=0$ by definition), required to maximize a figure of merit
which we refer to as the number of unique and correct associations.  This figure of merit measures the probability that a given simulated light
curve is correctly and uniquely associated with the family of models it belongs to, so as to ensure that the viewing angle ($\theta_v$) and
circumstellar density ($n_{\rm ISM}$) of the radio counterpart are correctly reconstructed from the observations. In other words, our goal is to
optimize the identification of the physical parameters of BNS radio counterparts, and to distinguish them from contaminants when these might
be relevant (see Section~\ref{sec:methods}). Our figure of merit is computed as follows.

For each of the 82$\times$4000 simulated observations of target light curves, we determine which models (both targets and contaminants)
predict fluxes $F_{\rm model}$ that, at $t_{\rm radio,0}$, agree with the simulated target observation i.e.,
$\left|F_{\rm model}-F_{\rm obs}\right|<3\times\sqrt{\sigma^2_{\rm model}+\sigma^2_{\rm obs}}$.
Models that satisfy this condition are considered positive associations for the first epoch $t_{\rm radio,0}$, and carried forward to the next
observing epoch.
For GW170817-like transients, we use the $\pm1\sigma_{\rm model}$ range on the structured jet best fit light curve as provided by \citet{Lazzati2017}.
For off-axis relativistic jets, we set $\sigma_{\rm model}=0$ but use current uncertainties in the microphysics parameters $\epsilon_E$ and
$\epsilon_B$ to reflect the related uncertainty on model predictions ($F_{\rm model}$).  Practically, this is implemented by allowing
$F_{\rm model}$ to have a scatter $\Delta F_{\rm \epsilon_E,\epsilon_B}$ determined by varying $\epsilon_E$ and $\epsilon_B$ in their
allowed ranges, while keeping all other model parameters unchanged. This is equivalent to considering all models differing only by their
values of $\epsilon_E$ and/or $\epsilon_B$ as part of the same target family. If at a certain epoch
$\left|F_{\rm family}-F_{\rm obs}\right|\lesssim 3\sigma_{\rm obs}$, for any value of $F_{\rm family}$ in the range
$F_{model}-\Delta F_{\rm \epsilon_E,\epsilon_B}\lesssim F_{\rm family}\lesssim F_{\rm model}+\Delta F_{\rm \epsilon_E,\epsilon_B}$,
then we call the above a correct association with the family. The association is also referred to as unique if only one single family agrees
with the observed flux.

After observing at $t_{\rm radio, 0}$, the second epoch can happen with any time delay, $\Delta T_1=M_1\times 2$\,d, with respect to the
first observation $t_{\rm radio, 0}$. In general, only a subset of the models or families that represented positive associations at $t_{\rm radio, 0}$
will also be positive associations for epoch two (i.e. will show agreement between observed flux and predicted model flux at that epoch within
$3\times\sqrt{\sigma^2_{\rm model}+\sigma^2_{\rm obs}}$). Thus, we optimize the value of $M_1$ by maximizing the number of associations
that in epoch two become unique (only one model or family fits the observed target in both epochs) and correct (the model or family that fits
the observations uniquely is also the correct one, i.e. it is the same target model or family from which the observations were simulated). 

We then add a third observing epoch, keeping the two already analyzed in place. As before, the third epoch can happen with any time delay, 
$\Delta t_2=M_2\times 2$\,d, with respect to the first observation. So we optimize $M_2$ by maximizing the number
of associations that, after being positive in both epoch one and two, become unique and correct associations in epoch three.

We keep repeating this process until we reach a maximum of ten epochs.
Naturally, adding more observational epochs will progressively increase the fraction of unique and
correct associations up to that epoch (see Table~\ref{tab:results}).
Note that it may happen that in the optimization process $M_n$ turns out to be larger than $M_{n+1}$. Thus, the times of the optimized
observational epochs are ordered in increasing delays since first epoch once the optimization process is completed (Table~\ref{tab:results}).

\begin{center}
\begin{table*}
\caption{Results summary for different observing strategies. The first observation is always assumed to be performed as soon as possible
(i.e. within the time interval determined by the level of urgency). The timing of subsequent epochs is set so as to maximize the number of unique
and correct associations. We show how the fraction of unique and correct associations grows by increasing the total number of observations,
up to the maximum of ten that was assumed in this analysis. 
\label{tab:results}}
\resizebox{\textwidth}{!}{
\begin{tabular}{c|l|c|l|c|l|c}
\hline
\hline
& \multicolumn{2}{c}{High urgency} & \multicolumn{2}{|c|}{Medium urgency} & \multicolumn{2}{c}{Low urgency} \\ 
\hline
\#of Epochs & days since 1$^{\rm st}$ obs. & unique \& correct &days since 1$^{\rm st}$ obs. & unique \& correct & days since 1$^{\rm st}$ obs. & unique \& correct \\
\hline
2 & 6 & 40.5\% & 6 & 39.5\% & 6 & 24.3\%\\
3 & 6, 54 & 47.6\% & 6, 42 & 46.2\% & 6, 38 & 31.5\%\\
4 & 6, 8, 54 & 53.7\% & 2, 6, 42 & 48.4\% & 2, 6, 38 & 36.3\%\\
5 & 2, 6, 8, 54 & 55.5\% & 2, 4, 6, 42 & 48.9\% & 2, 4, 6, 38 & 38.8\%\\
6 & 2, 4, 6, 8, 54 & 55.8\% & 2, 4, 6, 10, 42 & 49.7\% & 2, 4, 6, 8, 38 & 40.4\%\\
7 & 2, 4, 6, 8, 10, 54 & 58.6\% & 2, 4, 6, 8, 10, 42 & 53.1\% & 2, 4, 6, 8, 18, 38 & 42.6\%\\
8 & 2, 4, 6, 8, 10, 54, 62 & 59.3\% & 2, 4, 6, 8, 10, 42, 58 & 54.8\% & 2, 4, 6, 8, 18, 38, 62 & 45.0\%\\
9 & 2, 4, 6, 8, 10, 14, 54, 62 & 59.9\% & 2, 4, 6, 8, 10, 42, 50, 58 & 55.1\% & 2, 4, 6, 8, 18, 22, 38, 62 & 45.6\%\\
10 & 2, 4, 6, 8, 10, 14, 54, 58, 62 & 60.7\% & 2, 4, 6, 8, 10, 42, 50, 54, 58 & 55.6\% & 2, 4, 6, 8, 10, 18, 22, 38, 62 & 45.9\%\\
\hline
\end{tabular}
}
\end{table*}
\end{center}

\begin{table*}
\begin{center}
\caption{Results summary for different targets with a maximum of eight or ten total observing epochs (at the times specified in
Table~\ref{tab:results}). See text for discussion.}
\label{tab:results_targets}
\begin{tabular}{c|ccc|ccc}
\hline
\hline
& \multicolumn{3}{c}{8 epochs} & \multicolumn{3}{|c}{10 epochs} \\
& High urgency & Medium urgency & Low urgency & High urgency & Medium urgency & Low urgency \\ 
\hline
Class & unique \& correct  & unique \& correct & unique \& correct & unique \& correct& unique \& correct& unique \& correct\\
\hline
Target 0 & 	86.2\% & 71.8\% & 53.1\% & 	86.2\% & 71.8\% & 53.5\% \\
Target 1 & 	60.4\% & 59.0\% & 51.4\% & 	60.4\% & 59.0\% & 52.2\% \\
Target 2 & 	0\% & 0\% & 0\% & 			0\% & 0\% & 0\% \\
Target 3 & 	62.3\% & 59.0\% & 48.5\% & 	65.3\% & 59.0\% & 48.8\% \\
Target 4 & 	0\% & 0\% & 19.4\% & 		8.0\% & 7.9\% & 24.8\% \\
Target 5 & 	0\% & 0\% & 0\% & 			0\% & 0\% & 0\% \\
Target 6 & 	20.9\% & 30.2\% & 5.0\% & 	20.9\% & 30.2\% & 5.0\% \\ 
Target 7 & 	0\% & 0\% & 0\% & 			0\% & 0\% & 0\% \\
Target 8 & 	0\% & 0\% & 0\% & 			0\% & 0\% & 0\% \\
GW170817 & 	100\% & 100\%& 100\% & 	100\% & 100\% & 100\% \\
\hline
\end{tabular}
\end{center}
\end{table*}

\section{Results}\label{sec:results}
The results of our analysis are summarized in Tables~\ref{tab:results} and \ref{tab:results_targets}.
We note that target families 5, 7 and 8 all predict radio peak fluxes that at 40\,Mpc  are below the detection threshold of our simulated
observing campaign (see  Table~\ref{tab:dist}). Therefore these targets are never detectable nor identifiable by our simulations.
The fraction of unique and correct associations we quote in Tables~\ref{tab:results} and \ref{tab:results_targets} already take this into
account (i.e., we normalize these fractions by the total number of simulated targets that can at least in principle be detected
at least at peak given the assumed detection threshold of $3\times$\,RMS).
We also note that some models belonging to target family 2 are detected, but never uniquely identified. These sources are bright enough
to be detected only for a few days, and in that period their light curves are very similar to others. Thus, hereafter we focus on
the performance of the various observational strategies with respect to targets 0, 1, 3, 4, 6, and GW170817-like transients.

Using a high-urgency strategy, target 6 sources are uniquely associated only after seven observations, and target 4 sources are
uniquely associated only after ten observations (Table~\ref{tab:results_targets}). This is due to the fact that both families are barely detectable,
for the brightest combinations of $\epsilon_E$ and $\epsilon_B$ and, in both cases, the peak fluxes are easily confused with other models,
thus requiring several observations to break degeneracies.
After eight epochs, 86.2\% of the target 0 sources, about 60\% of target 1 and 3, and 20.9\% of target 6 are uniquely and correctly identified,
while all of the GW170817-like ones are. GW170817-like sources, in particular, are uniquely identified via late-time observations when
their light curves are distinctly different than the other types. Adding two more epochs allows us to correctly and uniquely associate up to 65.3\%
of target 3 sources, and up to 8.0\% of target 4 sources. For all other target families, adding more observations yields no further gain in correct
and unique associations.

A trend similar to the one described above is observed for the medium-urgency strategy. In this case, the correct and unique association 
percentages are lower for all targets but for target 6 sources (30.2\%), and for GW170817-like transients (that are still
correctly and uniquely associated in all cases).
After eight observations, 71.8\% of target 0, and 59.0\% of target 1 and target 3 sources are correctly and uniquely associated.
With two more observations up to 7.9\% of target 4 sources
are correctly and uniquely associated.

With a low-urgency strategy, after eight observations we are able to associate 19.4\% of target 4 sources, and only about 50\% of target 0, 1,
and 3 sources. However, the percentage of correct and unique associations for target 6 drops dramatically to only 5.0\%.
Adding two more epochs does not change much the correct ad unique association fractions for targets 0, 1, 3 and 6, while for target 4
we reach a value of 24.8\%.

In summary, most of the correct and unique associations for sources belonging to target families 0, 1, 3 and for GW170817-like transients
are obtained after seven epochs, for all priorities.
Eight epochs are necessary to maximize the correct and unique associations for target 6 sources. Ten epochs are needed to maximize
the correct and unique associations of target 4 sources. Targets 0, 1, 3 and GW170817-like sources are most efficiently identified with
a high-priority strategy, while a medium-priority is best for identifying target 6 sources. This difference in performance for target 6 is related
to the timing of the last two epochs in the eight-epoch observing strategy necessary for this target. 

In light of the above, we explore an ad-hoc strategy which consists of observing at 2, 4, 6, 8, and 10 days after the first epoch as determined
by the high-urgency strategy, plus two additional late-time epochs drawn by the results of the medium-urgency strategy (42\,d and 58\,d since
first epoch in medium urgency, that correspond to 45\,d and 61\,d after the first epoch in high urgency, respectively).
With this strategy (Table~\ref{tab:results_best}), the overall percentage of unique correct associations is 59.2\%. Compared with the high-urgency
strategy in Table~\ref{tab:results_targets}, the same amount of target 0, target 4, and GW170817-like sources are identified.
While 59\% of target 1 sources are identified (fewer than before), more correct associations are achieved for target 3 and target 6 sources.
Finally, this ad-hoc strategy would not allow us to identify any target 4 sources.

\subsection{Excluding contaminant sources}
We have also repeated our simulations removing all contaminant sources. This simulates the case when a clear, undisputed association between
the optical counterpart and the BNS event
has been established before the radio monitoring begins (as for GW170817).
We compare our simulated observations against a bank of radio light curves containing only possible BNS radio counterparts
(``targets'' in Table~\ref{tab:models}), and optimize the radio follow-up strategy as explained in Section~\ref{sec:targets} using the high-urgency
setup (see Section~\ref{sec:urgency}). Our results are listed in Table~\ref{tab:no_contaminants}.
As evident from this Table, removing contaminants modifies our results only slightly.
The optimal strategy in this case is to perform 8 observations 2, 6, 8, 10, 14, 34, and 70 days after the first one, and the overall
percentage of unique and correct association is 61.1\%.

\begin{center}
\begin{table*}
\caption{
Summary of the results obtained including or excluding contaminant sources.
These results are obtained using the high urgency setup as described in Section~\ref{sec:urgency}.
It is evident that excluding contaminants only slightly improves our results.
\label{tab:no_contaminants}}
\resizebox{\textwidth}{!}{
\begin{tabular}{c|l|c|l|c}
\hline
\hline
& \multicolumn{2}{c}{With contaminants} & \multicolumn{2}{|c}{Without contaminants} \\ 
\hline
\#of Epochs & days since 1$^{\rm st}$ obs. & unique \& correct & days since 1$^{\rm st}$ obs. & unique \& correct \\
\hline
2 	& 6	& 40.5\% & 6	& 40.8\% \\
3 	& 6, 54	& 47.6\% & 6, 70	& 47.6\% \\
4 	& 6, 8, 54	& 53.7\% & 6, 8, 70	& 54.1\% \\
5 	& 2, 6, 8, 54	& 55.5\% & 2, 6, 8, 70	& 55.8\% \\
6 	& 2, 4, 6, 8, 54	& 55.8\% & 2, 6, 8, 10, 70	& 56.3\% \\
7 	& 2, 4, 6, 8, 10, 54	& 58.6\% & 2, 6, 8, 10, 34, 70	& 59.4\% \\
8 	& 2, 4, 6, 8, 10, 54, 62	& 59.3\% & 2, 6, 8, 10, 14, 34, 70	& 61.1\% \\
9 	& 2, 4, 6, 8, 10, 14, 54, 62	& 59.9\% & 2, 4, 6, 8, 10, 14, 34, 70	& 61.2\% \\
10	& 2, 4, 6, 8, 10, 14, 54, 58, 62	& 60.7\% & 2, 4, 6, 8, 10, 14, 18, 34, 70	& 61.2\% \\
\hline
\end{tabular}
}
\end{table*}
\end{center}

\begin{table*}
\begin{center}
\caption{Results summary for different targets using the ad-hoc strategy discussed in Section~\ref{sec:conclusions}.
This strategy involves high urgency, and a total of 8 observations. If a source has been uniquely and correctly associated after
5 observations, then the remaining 6 epochs are not performed.
See text for discussion.}
\label{tab:results_best}
\begin{tabular}{c|c|c}
\hline
\hline
& \multicolumn{2}{c}{Percentage of unique \& correct associations} \\ 
& First 5 observations & 8 observations \\
Class & 2, 4, 6, and 8 days after the first epoch & 2, 4, 6, 8, 10, 45, and 61 days after the first epoch \\
\hline
Target 0 & 86.1\%	& 86.2\%	\\
Target 1 & 54.4\%	& 59.4\%	\\
Target 2 & 0\%		& 0\%	\\
Target 3 & 62.3\%	& 63.2\%	\\
Target 4 & 0\%		& 0\%	\\
Target 5 & 0\%		& 0\%	\\
Target 6 & 0\%		& 23.6\%	\\
Target 7 & 0\%		& 0\%	\\
Target 8 & 0\%		& 0\%	\\
GW170817 & 0\%	& 100\%	\\
\hline
\end{tabular}
\end{center}
\end{table*}

\begin{table}
\begin{center}
\caption{Maximum distance at which each type of target source can be detected using the sensitivity of the current JVLA and of ngVLA.
\label{tab:dist}}
\begin{tabular}{c|c|c}
\hline
\multirow{2}{*}{Class} & JVLA Distance & ngVLA Distance \\
 &  Horizon (Mpc) &  Horizon (Mpc) \\
\hline
Target 0 &	288 & 910 \\
Target 1 &	147 & 465 \\
Target 2 &	64 & 203 \\
Target 3 &	136 & 429 \\
Target 4 &	56 & 176 \\
Target 5 &	21 & 67 \\
Target 6 &	44 & 138 \\
Target 7 &	17 & 53 \\
Target 8 & 6 & 19 \\
\hline
\end{tabular}
\end{center}
\end{table}

\section{Follow-up at other wavelengths}
\label{sec:discussion}
To establish whether optical and/or X-ray follow-up observations can be used to probe emission from fast jets in BNS mergers,
using the BOXFIT code we have simulated light curves for all our target sources in $R$-band (658\,nm) and at 1\,keV.
We have then estimated their fluxes at $d_L=40$\,Mpc, and calculated the earliest and latest epochs at which each light curve model
would be detectable.

None of the optical afterglows of the models considered here are detectable as their maximum brightness would be above magnitude 24 in R-band.
For what concerns X-ray afterglows, with \textit{Swift} in a $\approx 10$\,ks-long observation one can reach a $3\sigma$ sensitivity of
$\sim2.5\times10^{-14}$\,erg\,cm$^{-2}$\,s$^{-1}$ (unabsorbed flux), and none of our models would be detectable.
With a 20\,ks-long observation with \textit{Chandra} one could reach a $3\sigma$ sensitivity of $\sim 3\times10^{-15}$\,erg\,cm$^{-2}$\,s$^{-1}$
(unabsorbed flux).
In this case, some models belonging to target 0 family
and the model with the highest values of $\epsilon_E$ and $\epsilon_B$ for target 3 family, reach an X-ray peak flux that is detectable.
All these models become detectable between 10 hours and 4 days after the merger, and remain detectable for a period of a few hours to a
couple of weeks.

\begin{figure}
\begin{center}
\includegraphics[width=0.49\textwidth,viewport=2 0 530 400,clip]{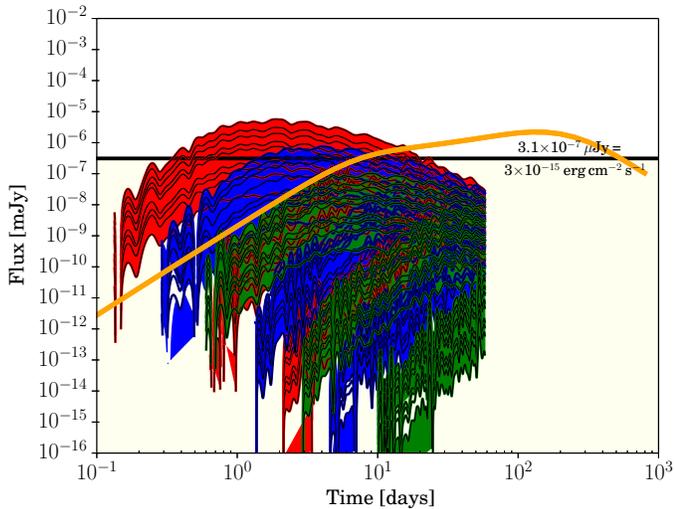}
\caption{
1\,keV light curves of our target sources, as well as the one for GW170817. The color coding is the same as in Figure~\ref{fig:models}.
}
\label{fig:models_Xrays}
\end{center}
\end{figure}

The most optimistic prospects for X-ray detections come from GW170817-like transients. Indeed, as shown in Figure~\ref{fig:models_Xrays},
the best fit structured jet model for GW170817 at 1\,keV \citep[model from ][]{Lazzati2017} predicts that the X-ray afterglow becomes bright
enough to be detected with \textit{Chandra} a few days after the merger, and remains visible for a long time (up to about 1000\,d).

We conclude that radio is critical to probing the dynamics of the relativistic jets, as optical and X-rays are often too faint to be detected.

\section{Discussion and Conclusion}\label{sec:conclusions}
We have presented an analysis aimed at identifying an optimal strategy for detecting and uniquely associating GHz off-axis radio afterglows 
and GW170817-like radio counterparts of nearby ($d_L\sim 40$\,Mpc) BNS mergers.

We established that eight 2\,h-long JVLA observations are needed to maximize the unique and correct associations for the possible BNS
radio counterparts considered in this work. The fraction unique and correct associations reach 86\% for target 0, 60\% for target 1, 62\%
for target 3, and 100\% for GW170817-like transients.
We have also determined that a high-urgency strategy in which the first JVLA follow-up observation is carried out within 2\,d since optical
identification is necessary for most targets.
One would then carry out radio follow-up observations at 2, 4, 6, and 8 days after the first radio epoch, enabling the identification
of the majority of target 0, 1, and 3 sources.
If after eight days since the first radio epoch the observed radio emission is still consistent with target 4, or target 6, or with a GW170817-like
transient, or if a unique and correct association with one of the other targets has not yet been established, three additional epochs at
10, 45, and 61 days since the first are desirable (see Table~\ref{tab:results_best}).
We have also shown that, in most cases, radio is the most effective at probing off-axis relativistic jets from BNS mergers expanding
in a low-density ISM, since their optical and X-ray counterparts are often too faint.

In Table~\ref{tab:dist} we show the maximum distance up to which the brightest model of each of the off-axis SGRB
families used here are detectable with a 2\,hr-long JVLA observation. As evident from this Table, target 0 (ISM density of
10$^{-2}$\,cm$^{-3}$ and $\theta_v=20$\,deg) is the only family with at least a model bright enough to be detectable at the distance horizon
of Advanced LIGO \citep[$\sim$200\,Mpc;][]{Abbott2016}. Thus, next generation radio arrays will play a key role in expanding the parameter
space accessible to radio follow ups of BNS mergers. Particularly, the next generation Very Large Array (ngVLA) will be $\sim10\times$
more sensitive than the current JVLA, tremendously improving our prospects for radio studies of BNS merger afterglows explored here.
To demonstrate this, we repeated our analysis simulating observations carried out with a radio array $10\times$ more sensitive than the JVLA
($3\sigma$ sensitivity of $\sim 1.5\,\mu$Jy) with sources $3\times$ as distant, ($d_L\approx 120$\,Mpc,).
Comparing the results of these simulations (Tables~\ref{tab:results_ngvla} and \ref{tab:results_targets_ngvla}) to the previous ones
(Tables~\ref{tab:results} and \ref{tab:results_targets}), we see that
the ngVLA will obtain the same results as the current generation JVLA for sources up to distances $3\times$ as large, for which the expected
rates are a factor of $\sim 30\times$ larger.

\begin{center}
\begin{table*}[!hbtp]
\caption{Results summary for different observing strategies for sources at 120\,Mpc, observed by the ngVLA, i.e. 3$\sigma$ sensitivity of
1.5\,$\mu$Jy.
\label{tab:results_ngvla}}
\resizebox{\textwidth}{!}{
\begin{tabular}{c|l|c|l|c|l|c}
\hline
\hline
& \multicolumn{2}{c}{High urgency} & \multicolumn{2}{|c|}{Medium urgency} & \multicolumn{2}{c}{Low urgency} \\ 
\hline
\#of Epochs & days since 1$^{\rm st}$ obs. & unique \& correct &days since 1$^{\rm st}$ obs. & unique \& correct & days since 1$^{\rm st}$ obs. & unique \& correct \\
\hline
2 & 8 & 42.1\% & 10 & 40.4\% & 6 & 25.7\%\\
3 & 8, 42 & 51.2\% & 10, 42 & 47.1\% & 6, 38 & 32.9\%\\
4 & 2, 8, 42 & 55.5\% & 2, 10, 42 & 50.0\% & 4, 6, 38 & 38.0\%\\
5 & 2, 6, 8, 42 & 56.0\% & 2, 4, 10, 42 & 50.5\% & 2, 4, 6, 38 & 40.8\%\\
6 & 2, 6, 8, 14, 42 & 56.7\% & 2, 4, 6, 10, 42 & 51.2\% & 2, 4, 6, 8, 38 & 41.8\%\\
7 & 2, 6, 8, 14, 42, 62 & 60.2\% & 2, 4, 6, 10, 42, 58 & 54.5\% & 2, 4, 6, 8, 38, 54 & 43.9\%\\
8 & 2, 4, 6, 8, 14, 42, 62 & 61.5\% & 2, 4, 6, 10, 14, 42, 58 & 56.7\% & 2, 4, 6, 8, 38, 54, 58 & 46.5\%\\
9 & 2, 4, 6, 8, 14, 42, 58, 62 & 62.2\% & 2, 4, 6, 10, 14, 42, 54, 58 & 59.7\% & 2, 4, 6, 8, 10, 38, 54, 58 & 47.0\%\\
10 & 2, 4, 6, 8, 10, 14, 42, 58, 62 & 63.6\% & 2, 4, 6, 10, 14, 18, 42, 54, 58 & 60.5\% & 2, 4, 6, 8, 10, 18, 38, 54, 58 & 47.8\%\\

\hline
\end{tabular}
}
\end{table*}
\end{center}

\begin{table*}[!hbtp]
\begin{center}
\caption{Results summary for different targets with a maximum of eight and ten total observing epochs (at the times specified in
Table~\ref{tab:results_ngvla}) for sources at 120\,Mpc, observed by the ngVLA. See text for discussion.}
\label{tab:results_targets_ngvla}
\begin{tabular}{c|ccc|ccc}
\hline
\hline
& \multicolumn{3}{c}{8 epochs} & \multicolumn{3}{|c}{10 epochs} \\
& High urgency & Medium urgency & Low urgency & High urgency & Medium urgency & Low urgency \\ 
\hline
Class & unique \& correct  & unique \& correct & unique \& correct & unique \& correct& unique \& correct& unique \& correct\\
\hline
Target 0 & 	86.9\% & 72.1\% & 52.9\% & 	86.9\% & 72.1\% & 53.2\% \\
Target 1 & 	60.8\% & 59.7\% & 49.6\% & 	62.6\% & 59.7\% & 52.3\% \\
Target 2 & 	0\% & 0\% & 0\% & 			0\% & 23.5\% & 4.2\% \\
Target 3 & 	66.5\% & 65.5\% & 46.5\% & 	66.7\% & 65.5\% & 47.1\% \\
Target 4 & 	0\% & 0\% & 13.5\% & 		16.9\% & 30.3\% & 21.3\% \\
Target 5 & 	0\% & 0\% & 0\% & 			0\% & 0\% & 0\% \\
Target 6 & 	33.3\% & 33.3\% & 30.7\% &	33.3\% & 33.3\% & 30.7\% \\ 
Target 7 & 	0\% & 0\% & 0\% & 			0\% & 0\% & 0\% \\
Target 8 & 	0\% & 0\% & 0\% & 			0\% & 0\% & 0\% \\
GW170817 & 	100\% & 100\%& 100\% & 	100\% & 100\% & 100\% \\
\hline
\end{tabular}
\end{center}
\end{table*}

\acknowledgments
D.C. and A.C. acknowledge support from the National Science Foundation CAREER Award \#1455090, and partial support from the Swift
Cycle 12 GI program (Grant  \# NNX17AF93G).  

\bibliographystyle{aasjournal}
\bibliography{./bibliography.bib}

\end{document}